\documentclass{elsart}

\usepackage{epsfig}

\begin{document}
\begin{frontmatter}

\title{Cosmochemical Consequences of Particle Trajectories During
FU Orionis Outbursts by the Early Sun}

\author{Alan P. Boss$^1$, Conel M. O'D. Alexander$^1$, Morris Podolak$^2$}
\address{$^1$Dept. of Terrestrial Magnetism, Carnegie Institution, 
5241 Broad Branch Road, NW, Washington, DC 20015-1305}
\address{$^2$Dept. of Geophysics \& Planetary Sciences, Tel Aviv 
University, Ramat, Tel Aviv, 69978 Israel}

Corresponding Author E-mail: boss@dtm.ciw.edu

Corresponding Author Telephone: 1-202-478-8858

\begin{abstract}

The solar nebula is thought to have undergone a number of episodes of FU Orionis outbursts during its early evolution. We present here the first calculations of the trajectories of particles in a marginally gravitationally unstable solar nebula during an FU Orionis outburst, which show that 0.1 to 10 cm-sized particles traverse radial distances of 10 AU or more, inward and outward, in less than 200 yrs, exposing the particles to temperatures from $\sim$ 60 K to $\sim$ 1500 K. Such trajectories can thus account for the discovery of refractory particles in comets. Refractory particles should acquire Wark-Lovering-like rims as they leave the highest temperature regions of the disk, and these rims should have significant variations in their stable oxygen isotope ratios. Particles are likely to be heavily modified or destroyed if they pass within 1 AU of the Sun, and so are only likely to survive if they formed in the final few FU Orionis outbursts, or were transported to the outer reaches of the solar system. Calcium, aluminum-rich inclusions (CAIs) from primitive meteorites are the oldest known solar system objects and have a very narrow age range. Most CAIs may have formed at the end of the FU Orionis outbursts phase, with an age range reflecting the period between the last few outbursts.

\end{abstract}


\end{frontmatter}

\section{Introduction}

Comets and asteroids are the leftover building blocks from the assembly of our solar system. Their isotopic and compositional properties are thus used to infer the processes that accompanied the formation and growth of the first solids in the solar protoplanetary disk (solar nebula). To date, the earliest-formed components of chondrites, the most primitive meteorites and fragments of asteroids, are the up to cm-sized CAIs. CAIs are thought to have formed during a brief period in the hot, innermost solar nebula. At what stage in the solar nebula's evolution CAIs formed is not known, although it is often assumed that they mark the start of solar system formation. Samples from Comet Wild 2 (Brownlee et al., 2006) contain abundant refractory grains, including CAIs. Volatile ice-rich comets must have formed in the cool outer solar system, requiring outward transport of refractory solids from the hot inner solar nebula against the inward flow of material onto the growing protosun. Recently, a CAI from the Allende meteorite was shown to retain evidence of repeated passages both inward and outward in the nebula (Simon et al., 2011a).

The two main competing mechanisms for large-scale transport of solids are ballistic trajectories above the disk and transport within the disk by hydrodynamical processes. The leading mechanism in the first case is the X-wind bipolar outflow model (Shu et al., 1997), but it has fallen into disfavor for a variety of reasons (Desch et al., 2010). Several distinct versions of the in-disk transport mechanism have been investigated. In the first, a generic turbulent viscosity is assumed to drive both the overall inward movement of material onto the protosun at high altitudes in the disk, and radially outward (meridional) flow at the disk midplane (Gail, 2002; Keller and Gail, 2004; Ciesla, 2007). The favored source of this turbulence is the magneto-rotational instability (MRI) (Hawley, Balbus, and Winters, 1999). However, the MRI instability is stifled in the magnetically dead disk midplane (Matsumura and Pudritz, 2006), potentially halting outward transport of solids driven by MRI turbulent diffusivity. In fact, such midplane dead zones are likely to accumulate enough mass to become gravitationally unstable (Zhu et al., 2011; but see also Terquem, 2008). The recent discovery that MRI disks do not in fact result in meridional circulation currents capable of large-scale transport (Fromang, Lyra, and Masset, 2011) casts further doubt on generic turbulent viscosity disk models.

Large-scale hydrodynamic transport could also be achieved by the outward expansion of an initially small viscous disk (e.g., Ciesla, 2010a;
Jacquet, Fromang, and Gounelle, 2011), though an MRI source for the viscosity
encounters the problems noted above. Turbulent diffusion of particles,
caused by a generic turbulent viscosity, has also been advanced as
a transport mechanism (e.g., Cuzzi, Davis, and Dobrovolskis, 2003;
Ciesla, 2010b). Charnoz et al. (2011) have shown that such diffusive
transport appears to operate in the opposite sense of the meridional
circulation currents previously hypothesized. Such diffusive transport
depends on the existence of some generic source of turbulence.

The remaining hydrodynamic mechanism is transport driven by spiral arms in gravitationally unstable regions of the disk. The formation of the solar systems gas giant planets appears to require that the solar nebula was massive enough at some phase to become at least marginally gravitationally unstable (MGU). The core accretion models of Inaba, Wetherill, and Ikoma (2003) and Chambers (2008) required disk masses of 0.08 $M_\odot$ and 0.1 $M_\odot$, respectively, while Alibert et al. (2005) presented results for a disk mass of 0.05 $M_\odot$. The alternative formation mechanism, disk instability (Boss, 1997), seems to require a disk mass of at least 0.04 $M_\odot$ to operate (Boss, 2010). Disks with masses in this range are likely to be MGU disks. 

Three-dimensional (3D) models have shown that MGU disks are capable of the rapid radial transport of gas and small particles both inward and outward (Boss, 2007; Boss, 2008), as well as the rapid homogenization of isotopic spatial heterogeneity (Boss et al., 2010; Boss and Keiser, 2010) over spatial scales ranging from a fraction of an AU to 20 AU (Boss, 2011). In addition, MGU disks are likely to be the source of the FU Orionis outbursts experienced periodically by all solar-type protostars (e.g., Zhu, Hartmann, and Gammie, 2010; Vorobyov and Basu, 2010). The previously cited problems with MRI disk models apparently leave a MGU disk phase as the leading candidate mechanism for achieving the global transport associated with FU Orionis events.

Our previous models (Boss, 2007; Boss, 2008) of disk transport used a color field equation, similar to the hydrodynamical continuity equation, to follow the trajectories of gas particles and dust grains small enough to remain coupled to the gas. Here we present the first results of MGU disk models where the particles are represented as finite-size bodies that are subject to gas drag (Haghighipour and Boss, 2003) and the gravity of the protosun. This allows the trajectories of individual particles to be followed in time, and predictions made of the thermal and chemical processing they would experience.

\section{Numerical Hydrodynamics Methods}

 The disk evolution calculations were performed with a numerical code
that uses finite differences to solve the three-dimensional equations 
of hydrodynamics, radiative transfer, and the Poisson equation for the
gravitational potential. The code is the same as that used in many
previous studies of mixing and transport in disks 
(e.g., Boss, 2007; Boss, 2008). The code has been shown to
be second-order-accurate in both space and time through convergence testing
(Boss and Myhill 1992). The equations are solved on a spherical coordinate
grid. The number of grid points in each spatial direction is: $N_r = 51$,
$N_\theta = 23$ in $\pi/2 \ge \theta \ge 0$, and $N_\phi = 256$. This
relatively low degree of numerical spatial resolution was chosen
in order to evolve the disks as far forward in time as possible 
in several years of computing on a dedicated workstation. The radial 
grid is uniformly spaced ($\Delta r = 0.18$ AU) between 1 and 10 AU, 
with boundary conditions at both the inner and outer edges chosen to 
absorb radial velocity perturbations and so to mimic a solar nebula
that extends from the central protostar to well beyond 10 AU. 
The $\theta$ grid is compressed into the midplane to ensure
adequate vertical resolution ($\Delta \theta = 0.3^o$ at the midplane).
The $\phi$ grid is uniformly spaced, to prevent any bias in the azimuthal
direction. The central protostar increases slightly in mass during the 
evolutions as a result of the accretion of disk gas, 
and wobbles in response to the growth of
nonaxisymmetry in the disk, thereby rigorously preserving the location of
the center of mass of the star and disk system. The number of terms in the
spherical harmonic expansion for the gravitational potential of the disk
is $N_{Ylm} = 32$. The four hydrodynamic equations are solved explicitly in 
conservation law form; artificial viscosity is not employed. 

\section{Initial Conditions for the Disk}

 The disk model consists of a $1 M_\odot$ central protostar surrounded
by a protoplanetary disk with a mass of 0.047 $M_\odot$ between 1 and
10 AU (e.g., Boss, 2008). Estimates of the minimum disk mass necessary
to form our solar system range from 0.01 $M_\odot$ to 0.1 $M_\odot$
(Weidenschilling, 1977), so this initial disk mass seems reasonable
on that basis alone. The underlying disk structure is the 
same as that of the disk extending from 4 to 20 AU assumed in previous 
models (e.g., Boss, 2007). 
Initially the disk has the density distribution (Boss, 1993) of
an adiabatic, self-gravitating, thick disk in near-Keplerian rotation 
about a stellar mass $M_s$

\begin{equation}
\rho(R,Z)^{\gamma-1} = \rho_o(R)^{\gamma-1} - \biggl( 
{ \gamma - 1 \over \gamma } \biggr) 
\biggl[\biggl( { 2 \pi G \sigma(R) \over K } \biggr) Z + 
{ G M_s \over K } \biggl( { 1 \over R } - { 1 \over (R^2 + Z^2)^{1/2} }
\biggr ) \biggr], 
\end{equation}

where $R$ and $Z$ are cylindrical coordinates,
$\rho_o(R)$ is the midplane density, and $\sigma(R)$ is the
surface density. The equations are solved for the top half of
a spherical volume ($Z > 0$), with the bottom half assumed to be
symmetrical with the top half. The adiabatic constant is $K = 1.7 \times 10^{17}$ 
(cgs units) and the adiabatic index $\gamma = 5/3$ for the initial model.
The radial variation of the initial midplane density is a power law 
that ensures near-Keplerian rotation throughout the disk

\begin{equation}
\rho_o(R) = \rho_{o4} \biggl( {R_4 \over R} \biggr)^{3/2},
\end{equation}

where $\rho_{o4} = (M_s/M_\odot) \times 10^{-10}$ g cm$^{-3}$ and
$R_4 = 4$ AU. Recent observations of low- and intermediate-mass pre-main-sequence 
stars imply disk masses in the range of 0.05-0.4 $M_\odot$ (Isella, Carpenter, and Sargent, 2009).

 The disk evolution starts from an axisymmetric density distribution with 
density perturbations of 1\% in the $m = 1$, 2, 3, and 4 modes. While the disk 
density distribution evolves in all four dimensions of space and time, in
order to speed up the calculations the disk 
evolves with a fixed, axisymmetric temperature distribution, with an outer 
disk temperature $T_o = 60$ K, rising to temperatures greater than
1500 K in the inner disk, based on the 
disk midplane temperature profiles calculated by Boss (1996).
This initial temperature distribution leads to a minimum in the 
initial Toomre $Q$ value of 1.8 at the outer boundary of the
active disk at 10 AU. Inside $\sim$ 5 AU, $Q$ rises to 
values $>$ 10 because of the much higher disk temperatures closer 
to the protosun. A $Q$ value of $>$ 1.5 implies marginal instability 
to the growth of gravitationally-driven perturbations, while $Q > 10$ 
implies a high degree of stability. In spite of the high inner $Q$ value,
the disk becomes strongly axisymmetric in the inner disk because of the
gravitational forces associated with the strong spiral arms and clumps
in the relatively low $Q$ region beyond $\sim$ 5 AU, as seen in 
Figure 1A after 105 yrs of evolution. It is only after this first
phase of 105 yrs that the particles are inserted into the disk and
their trajectories calculated, in order to speed the calculation
of the subsequent evolution. Note that by assuming an axisymmetric
temperature distribution, the particles will not be subject to thermal
processing associated with passage through spiral arm shocks, which is
likely to be another important physical effect (e.g., Wood, 1996; Boss and 
Durisen, 2005; Podolak, Mayer, and Quinn, 2011) for processing primitive 
materials. Here we focus instead on processing caused by the radial 
temperature gradient. 

\section{Gas Drag Effects on Particles}

 Two hundred particles, 1-cm-radius spheres with densities of 3 g cm$^{-3}$,
were followed for each model. The particles are represented as finite-size 
bodies that interact with the disk gas by gas drag, while 
moving on Keplerian orbits, subject to the gravity of the protosun 
and of the simultaneously evolving disk. 
The particles are not allowed to grow or fragment 
by collisions with other solids, and for simplicity are restricted to 
orbits in the disk midplane. In a more detailed calculation, vertical
motions should also occur, but given the recent indications that large-scale
meridional circulation currents are not likely to occur 
(Fromang, Lyra, and Masset, 2011), we believe that the radial motions
modeled here will dominate the large-scale particle transport.
 
 Gas drag effects are approximated using 
the expressions given by Haghighipour and Boss (2003), as follows. 

The gas drag force $F_d$ on a particle of radius $r_p$ in a gas
with a mean free path $\lambda$ is given by

\begin{equation}
\vec F_d = (1-f) \vec F_E + f \vec F_S,
\end{equation}

where $f = r_p / (r_p + \lambda)$, $\vec F_E$ is the Epstein drag law,
and $\vec F_S$ is the Stokes drag law. For gas composed of molecular
hydrogen, $\lambda = 4.72 \times 10^{-9}/ \rho$ cm, where $\rho$
is the gas density in cgs units. The Epstein and Stokes drag 
laws are given by

\begin{equation}
\vec F_E = - {4 \pi \over 3} r_p^2 \rho v_{th} \vec v_{rel},
\end{equation}

\begin{equation}
\vec F_S = - {0.44 \pi \over 2} r_p^2 \rho |v_{rel}| \vec v_{rel},
\end{equation}

where $v_{th}$ is the thermal velocity of the gas and $\vec v_{rel}$
is the relative velocity between the particle velocity $\vec v_p$ 
and the gas velocity $\vec v_g$

\begin{equation}
\vec v_{rel} = \vec v_p - \vec v_g.
\end{equation}

The Epstein drag law is appropriate for particles with $r_p << \lambda$,
while the Stokes law is appropriate for particles with $r_p >> \lambda$.
For particles with $r_p =$ 1 cm, both laws are applicable, depending on
the density of the region of the nebula (e.g., Figure 3D)
the particle resides in at the moment the gas drag is calculated.
When $|v_{rel}| << v_{th}$, the gas drag force becomes effectively
$\vec F_d = (1-f) \vec F_E$.

\section{Numerical Model for Thermal Processing of Dust Grain Particles}

Consider a system of solar composition with mass fractions $X_H$ of H$_2$, $X_{He}$ of He, $X_{H_2O}$ of water, $X_{CAI}$ of CAI-like material.  We assume that the gas and grains form a closed system, so that although mass can be transferred from one phase to another, no mass is lost from the system. In many CAIs, melilite is the dominant mineral. For simplicity in these order of magnitude calculations, we will assume that the CAI-like material is composed solely of melilite. The CAI cores are assumed to initially have radii of 0.5 cm, and all of the water in the system is assumed to be condensed upon them as an outer shell.  The total radius of the ice coated grains is initially very close to 1\,cm (and depends on the exact value chosen for the density of the rock material). The partial pressures of the condensable components are all assumed to be zero initially.

We assume mass fractions $X_H=0.7411$, $X_{He}=0.2457$, $X_{H_2O}=8.47\times 10^{-3}$, and $X_{CAI}=4.74\times 10^{-3}$.  We further assume that because the condensation temperature of water is so much lower than that of the rocky material, we can consider the phase changes of each species separately.  When the temperature is high enough for significant melilite evaporation, all of the water is already in the gas phase.  Conversely, when the temperature is low enough for water to condense, all of the melilite is already in the solid phase.  It should be noted that, in reality, because the kinetics of melilite condensation is so slow, due to the relatively low total pressures and surface areas of the CAIs, there will be a very small amount of melilite condensation even at the lowest temperatures, but we have ignored this small effect in our model.

\section{Calculation of Water Evaporation and Condensation}

The evaporation of water is computed via the expression

\begin{equation}
F=\left[P_{vap}^{wat}(T)-P_{wat}(T)\right]\sqrt{\frac{m_{\rm H_2O}}{2\pi kT}}
\label{evapflx}
\end{equation}

where $F$ is the mass flux of water evaporating, $P_{wat}$ is the partial pressure of water vapor in the surrounding gas (initially set to zero), $P_{wat}^{vap}$ is the equilibrium vapor pressure of water, $m_{\rm H_2O}$ is the mass of a water molecule, and $k$ is Boltzmann's constant.
  
A good fit to measurements of the equilibrium vapor pressure of water in the temperature range of interest is given by

\begin{equation}
P_{vap}^{wat}(T)=e^{-5640.23/T+22.867}\ \ \ {\rm bar}.
\label{pvapw}
\end{equation}

In a closed system the ratio of water to hydrogen is fixed.  The mass  
of water in the grains can be found from their ice shell thicknesses.   
The remainder of the water will be in the form of vapor, and this  
gives the value of $P_{wat}$.  The sign of $F$ in Equation 1 fixes  
whether the water is evaporating or condensing.  

\section{Calculation of Melilite Evaporation and Condensation}

The evaporation of minerals, such as melilite, is likely to be complex, and to date there have been no experimental measurements of the rate controlling mechanisms. Hence, for the purposes of our first order estimates of the extent of evaporation a melilite-dominated CAI would experience as it makes a passage close to the inner boundary of our MGU disk model, we only consider the thermodynamically most favorable evaporation and condensation reactions.

At equilibrium, evaporation and condensation rates are equal. Therefore, the maximum possible evaporation rate is controlled by the gaseous component in the equilibrium vapor that has, after accounting for the stoichiometry of the reaction, the slowest encounter frequency with the surface. This is usually the most massive component. We have considered a number of possible pathways for evaporation of gehlenite and akermanite, the end members of the melilite solid solution. The reactions used are the most thermodynamically favorable (i.e., the fastest reactions for the specific mechanisms under consideration here) under a given set of disk temperature and pressure conditions. 

One possibility is that gehlenite evaporates according to the reaction:\\

\centerline{Ca$_2$Al$_2$SiO$_7\Longleftrightarrow 2$Ca + 2AlO + SiO + 2O$_2$\ \ \ \ \ (R1)\\}

For this we have

\begin{equation}
-\frac{\Delta G}{RT}=\ln(P_{Ca}^2P_{AlO}^2P_{SiO}P_{O_2}^2)
\end{equation}

where $\Delta G$ is the change in the Gibbs free energy of the reaction, $T$ is the temperature, $R$ is the gas constant, and $P_x$ is the partial pressure of species $x$, now given in bars. Here the change 
in the Gibbs free energies of the reactions were calculated using the thermodynamic parameters given in previous compilations for minerals (Berman, 1988) and gases (Knacke, Kubachewski, and Hesselmann, 1991). Since each molecule of gehlenite that evaporates stoichiometrically releases one molecule of SiO, two molecules of Ca, etc. to the gas phase, then, assuming that the gas is ideal, at equilibrium the partial pressures of the various 
gases are in the proportions

\begin{equation}
2P_{Ca}=P_{SiO}=2P_{AlO}=2P_{O_2}.
\end{equation}

As the most massive molecule, SiO is likely to control the evaporation rate. Rewriting Eqn. 9 in terms of SiO alone gives

\begin{equation}
-\frac{\Delta G}{RT}=\ln(64 P_{SiO}^7),
\end{equation}

which can be rewritten as

\begin{equation}
P_{SiO}=\left(\frac{e^{-\Delta G/RT}}{64}\right)^{1/7}\ \ \ \ \ {\rm bar}.
\end{equation}

Corundum is a very refractory mineral. At temperatures below the condensation 
temperature ($\sim$ 1750 K) for corundum (Ebel, 2006, and references therein), 
a second more thermodynamically favorable pathway for gehlenite evaporation 
is\\

\centerline{Ca$_2$Al$_2$SiO$_7\Longleftrightarrow 2$Ca + Al$_2$O$_3$ (solid) + SiO + 1.5O$_2$\ \ \ \ \ (R2)\\}
for which

\begin{equation}
P_{SiO}=\left(0.13608e^{-\Delta G/RT}\right)^{1/4.5}\ \ \ \ \ {\rm bar}.
\end{equation}

Equilibrium condensation calculations for a solar gas predict that corundum would be replaced by hibonite (CaAl$_{12}$O$_{19}$), which in turn would be replaced by grossite (CaAl$_4$O$_7$), as the temperature falls
(Ebel, 2006, and references therein). 
The Ca/Al ratios of hibonite and grossite are much lower than that of gehlenite. Consequently, even if hibonite or grossite are slightly more stable than corundum under certain conditions, most of the Ca must still evaporate. For simplicity, only corundum is considered here.

For akermanite, the thermodynamically most favorable evaporation reaction is\\

\centerline{Ca$_2$MgSi$_2$O$_7\Longleftrightarrow 2$Ca + Mg + 2SiO + 2.5O$_2$,\ \ \ \ \ (R3)\\}

for which

\begin{equation}
P_{SiO}=\left(1.14487e^{-\Delta G/RT}\right)^{1/7.5}\ \ \ \ \ {\rm bar}.
\end{equation}

Each of these reactions is modified if H$_2$ gas is present.  R1 becomes\\

\centerline{Ca$_2$Al$_2$SiO$_7$ + 4H$_2\Longleftrightarrow 2$Ca + 2AlO + SiO + 4H$_2$O.\ \ \ \ \ (R$1'$)\\}

This gives

\begin{equation}
P_{SiO}=\left(2.4414\times 10^{-4} P_{H_2}^4e^{-\Delta G/RT}\right)^{1/9}\ \ \ \ \ {\rm bar}.
\end{equation}

With H$_2$, R2 becomes\\

\centerline{Ca$_2$Al$_2$SiO$_7$ + 3H$_2\Longleftrightarrow 2$Ca + Al$_2$O$_3$ (solid) + SiO + 3H$_2$O.\ \ \ \ \ (R$2'$)\\}

This gives

\begin{equation}
P_{SiO}=\left(9.2593\times 10^{-3} P_{H_2}^3e^{-\Delta G/RT}\right)^{1/6}\ \ \ \ \ {\rm bar}.
\end{equation}

Allowing for H$_2$ in reaction R3 gives\\

\centerline{Ca$_2$MgSi$_2$O$_7$ + 5H$_2\Longleftrightarrow 2$Ca + Mg + 2SiO + 5H$_2$O,\ \ \ \ \ (R$3'$)\\}

so we have

\begin{equation}
P_{SiO}=\left(0.20239 P_{H_2}^5e^{-\Delta G/RT}\right)^{1/10}\ \ \ \ \ {\rm bar}.
\end{equation}

If we do not allow for the reaction with H$_2$, then the mass flux of SiO coming off the surface of the grain into vacuum will be

\begin{equation}
F_{SiO}=\alpha P_{vap}^{SiO}\sqrt{\frac{m_{SiO}}{2\pi kT}},
\end{equation}

where$P_{vap}^{SiO}$ is the equilibrium vapor pressure of SiO,  $\alpha$ is the evaporation coefficient, $k$ is Boltzmann's constant, and 
$m_{SiO}$ is the mass of a SiO molecule. If there is some SiO already in the 
gas phase, then the net flux from the surface becomes

\begin{equation}
F_{SiO}=\alpha (P_{vap}^{SiO}-P_{gas}^{SiO})\sqrt{\frac{m_{SiO}}{2\pi kT}}.
\end{equation}

Here too the assumption of a closed system keeps the mass ratio of  
silicates to hydrogen constant.  Once the mass of silicates in the  
grains is known, the mass in the gas phase can be computed and the  
value of $P_{gas}^{SiO}$ calculated.

Since for stoichiometric evaporation of akermanite, for instance, every two molecules of SiO that leave the surface requires that a molecule of Ca$_2$MgSi$_2$O$_7$ has decomposed, we have

\begin{equation}
\frac{dM_{grain}}{dt}=-4\pi {r_p}^2\frac{F_{SiO}}{2}\frac{\mu_{akermanite}}{\mu_{SiO}}, 
\end{equation}

where $r_p$ is the radius of the particle and $\mu$ is the mean molecular weight ($\mu_{akermanite}=272.65$, $\mu_{gehlenite}=274.21$, $\mu_{Al_2O_3}=101.96$, and $\mu_{SiO}=44.09$).  So 

\begin{equation}
\frac{1}{4\pi {r_p}^2}\frac{dM_{grain}}{dt}=-8.998\times 10^3\alpha\frac{P_{vap}^{SiO}-P_{gas}^{SiO}}{\sqrt{T}}\ \ \ \ {\rm kg\,m^{-2}\,s^{-1}}
\label{newR2}
\end{equation}

If we use R$2'$, then each SiO molecule that evaporates causes one molecule of gehlenite to disappear and leaves behind a molecule of Al$_2$O$_3$.  The above equation
must now be replaced by

\begin{equation}
\frac{1}{4\pi {r_p}^2}\frac{dM_{grain}}{dt}=F_{SiO}\frac{\mu_{gehlenite}-\mu_{Al_2O_3}}{\mu_{SiO}}=-1.137\times 10^4\alpha\frac{P^{SiO}_{vap}-P^{SiO}_{gas}}{\sqrt{T}}
\ \ \ \ {\rm kg\,m^{-2}\,s^{-1}}
\label{newR3'}
\end{equation}

and with $P_{vap}^{SiO}$ appropriate to this reaction.  

We assume a temperature independent value for $\alpha$ for melilite of 0.1, which is of the order of values for forsterite and oxides evaporating from silicate melts (Alexandar, 2002;
Hashimoto, 1990; Kuroda and Hashimoto, 2002; Nagahara and Ozawa, 1996; Richter et al., 2007; Wang et al., 1999). The change in total pressure and temperature with time used in the above equations are those given 
by the disk transport calculations described above, where for each particle, we store the disk temperatures, pressures, and densities encountered (e.g., Figure 3) during the particle trajectories, for
use in post-processing with the thermal processing code.
We also assume that the partial pressure of H$_2$ is not modified by the evaporation and condensation reactions, or the dissociation to atomic H, and that the evaporation and condensation reactions do not appreciably modify 
the total pressure of the system. These are reasonable approximations given that H$_2$ is by far the most abundant component in a solar composition system under the conditions assumed here, and along with He essentially determines the total pressure of the system - at equilibrium at 1500 K and a total pressure of $10^{-5}$ bars, equilibrium between H and H$_2$ requires a H/H$_2$ ratio 
of $\sim$ 0.006.

While we believe that our present results are reasonably robust, we note that we have made several key simplifying assumptions, the effects of which should be examined in future studies. For example, we do not take into account in our calculations the following:
(1) the melilite solid solution, (2) a large fraction of the Mg and Si in a solar composition system would not be in CAIs and is not included in our calculations, (3) the potential reactions at the CAI surface that make new minerals, particularly during cooling, which may affect how the layering in WL rims formed, or (4) the possibility that  
when the corundum layer produced by evaporation of gehlenite is thick enough, it will protect the CAI against further evaporation until temperatures get high enough to start melting the entire CAI. 

We note that in a system with net evaporative losses, evaporation is by definition a non-equilibrium process - at equilibrium, there are no net losses. The conditions we consider do not even reach the solidus temperatures for CAIs, let alone the liquidus. For ablation to occur, e.g., one needs a large difference in velocity between gas and particle, as in atmospheric entry or shock heating. In our simulations, such large differences in velocity do not exist.

\section{Disk Model Results}

Several new sets of 3D MGU disk models including particles with radii ranging from 1 mm to 1 m have been computed for disks, where the particles are allowed to evolve within regions extending from 1 to 10 AU in a disk with a mass of 0.047 $M_\odot$. Two hundred particles with densities of 3 g cm$^{-3}$ were followed for each model. The particles are not allowed to grow or fragment by collisions with other solids, and for simplicity are restricted to orbits in the disk midplane. In order to speed the disk evolution calculations, the initial disk temperature distributions are held fixed, while the disk gas evolves, allowing the models to be followed for $\sim$ 200 yrs, i.e., 200 orbital rotation periods at the inner edge of the disk. 

Figure 1 shows the midplane density and temperature distributions in the disk at the beginning of the cm-sized particle evolutions and 205 yrs later. The particles began their evolutions within a disk that had already evolved for 105 yrs and developed spiral arms and non-axisymmetric structure (Fig. 1A). The disk became even more strongly non-axisymmetric 205 yrs later (Fig. 1B), with significant disk mass having been accreted by the central protostar, leaving behind low density regions in the inner disk, as well as significant outward transport of disk gas, indicated by the dense regions along the disks outer boundary. 

Our models are not meant to be detailed models of FU Orionis outbursts, but a comparison with astronomical observations of FU Orionis outbursts is appropriate here. Most solar-type protostars are thought to experience at least one FU Orionis outburst, based on the limited statistics of this phenomenon (e.g., Hartmann and Kenyon, 1996). In our model,
the central protostar gained 0.01 $M_\odot$ during the 205 yrs of evolution, yielding a central mass accretion rate of $5 \times 10^{-5} M_\odot$ yr$^{-1}$, comparable to rates estimated for $\sim$ 100-year-long FU Orionis outbursts occurring every $\sim 10^4$ yrs for T Tauri stars (Hartmann and Kenyon, 1996). In fact, a leading explanation for FU Orionis outbursts is a MGU disk (e.g., Zhu, Hartmann, and Gammie, 2010; Vorobyov and Basu, 2010). In spite of the disk mass loss in our model, the outer disk remains MGU and continues
to drive mixing and transport in the inner disk, though over a time scale of $\sim$ 200 yrs the mean central mass accretion rate drops by a factor of $\sim$ 100 (c.f., Figure 18 of Boss, 2011) and disk transport rates will be throttled back accordingly. A quiescent phase of disk evolution is expected in between MGU phases and FU Orionis outbursts. Presumably MRI effects in the regions of the disk beyond $\sim$ 10 AU, where cosmic ray ionization becomes important, are able to drive the inward transport of new disk mass to the dead zone inside $\sim$ 10 AU during relatively quiescent phases (e.g., Terquem, 2008). Once the disk again becomes MGU, another FU Orionis outburst can occur, apparently on a time scale of $\sim 10^4$ yrs.

The models show that particles with radii of 1 mm, 1 cm, or 10 cm that are initially close together can remain close together for hundreds of yrs, exhibiting a limited form of closed system behavior over this size range, while clumps of 1-m-sized bodies, for which gas drag effects peak (Haghighipour and Boss, 2003), tend to disperse on these time scales. Individual particles can move inward and outward in the disk repeatedly on eccentric orbits, sampling a wide range of nebular conditions. Here we focus on cm-size particles, with 0.5-cm-size refractory cores, typical
of the sizes of the best studied CAIs, although our results will hold for 1 mm to 10 cm particles as well.

Figure 2 displays the orbital evolutions of 200 particles of radius 1 cm that were uniformly placed in the disk midplane as shown in Figure 1A. While the particles start out moving in unison, after $\sim$ 20 yrs their trajectories begin to diverge significantly, with some first striking the inner disk boundary at 1 AU and being forced to remain there, while others later begin to reach the outer disk boundary at 10 AU. All of the particles strike and are held fixed at one or the other boundary within $\sim$ 200 yrs, with 72.5\% hitting the inner boundary and 27.5\% reaching the outer boundary. The oscillatory, yet somewhat chaotic, appearance of the trajectories is caused by the particles interactions with the MGU disks spiral arms, which are constantly evolving in amplitude and phase, and interacting with each other, both constructively and destructively. 

While a large fraction of the particles strike the inner boundary and are then removed from the calculation, that does not necessarily mean that a similarly large fraction would be lost to the growing protosun. However, the high temperatures (1550 K and above) inside 1 AU would be sufficient to partially melt and evaporate silicate grains, even during short visits to the inner disk, resetting their internal isochrons, and producing molten objects like Type B CAIs. Post-formational heating over a time period of $\sim$ 10 yrs has been inferred for the sublimation of at least one CAI (Shahar and Young, 2007). Equally, the fraction of particles that hit the outer boundary are not necessarily preserved. Further work will be needed to determine what fraction of CAIs during a MGU phase are: (a) likely to be altered yet survive, (b) completely evaporated or lost to the growing protosun, and (c) transported outward to comet-forming distances. 

The frequency with which CAIs encounter the inner boundary in our simplified models suggests that only a few FU Orionis events could rework or destroy most CAIs in a young, MGU-unstable disk. If correct, this leads to several far-reaching implications, given that a typical young stellar object will go through multiple FU Orionis events: (1) To explain the abundances of CAIs in many chondrites, and to compensate for those that are destroyed during each MGU episode, there must be a continuous mechanism for making CAIs in the nebula during the relatively quiescent inter-MGU phase, because grains traverse the CAI minerals stability region too quickly during MGU events to grow to the observed CAI sizes. (2) The majority of CAIs found in chondrites, with canonical $^{26}$Al/$^{27}$Al ratios, probably formed in the last FU Orionis event that the young solar system experienced, rather than at its formation, as is commonly assumed, naturally leading to the observed narrow range of CAI Al-Mg ages (MacPherson et al. 1995; Jacobsen et al. 2008; Larsen et al., 2011). (3) The much rarer evidence for supra-canonical 
$^{26}$Al/$^{27}$Al ratios (Young et al., 2005; Simon and Young, 2011) might
then derive from CAIs that were thoroughly thermally processed during an earlier 
FU Orionis event.

Figure 3 exhibits the journey of a representative 1 cm particle that evolved for 205 yrs before hitting the outer disk boundary at 10 AU. Starting at an orbital radius of 5.2 AU, the particle is first transported inward to close to 1 AU, but avoids the boundary. The particle is then transported outward, with several radial oscillations, before striking the outer disk boundary. This particle experienced ambient disk temperatures ranging from 60 K to a maximum of 1550 K. At least 10\% of the 1 cm particles that start out around 5 AU follow trajectories more or less similar to that of the particle in Figure 3. 

Trajectories such as that shown in Figure 3 are capable of delivering thermally processed, refractory particles to the comet-forming region of the solar nebula, even in the presence of massive clumps that might go on to form Jupiter (Figure 1A; Boss 1997), as is required to explain the refractory particles found in Comet Wild 2 (Brownlee et al., 2006). Comet Wild 2 particles have sizes of order $\sim$ 0.01 mm (e.g., Matzel et al., 2010; Ogliore et al., 2012), small enough to be transported with the gas, as studied by the color equation models of Boss (2007; 2008; 2011), but considerably smaller than the cm-sized particles shown here. Our models have shown that mm-sized particles have trajectories quite similar to those of cm-sized particles, and we expect the same to be true for 0.01 mm-sized particles: to first order, they will move with the gas, as in the color equation models. However, once the disk has lost enough mass to the protosun to no longer be marginally gravitationally unstable, Jupiter is expected to open a gap in the now relatively quiescent disk, possibly stopping altogether the flux of particles between the inner and outermost disk (e.g., Ogliore et al., 2012). However, once a gap opens around a Jupiter-mass protoplanet, Type III migration is thought to occur (e.g., Pepli\'nski, Artymowicz, and Mellema, 2008; Lin and Papaloizou, 2010), where disk gas continues to flow through the corotation gap, partially on horseshoe or tadpole orbits, and partially through the circumplanetary disk, leading to a reduced, but still significant, flux of disk gas and dust across the gap. This means that even the late-forming chondrules (Ogliore et al., 2012) and CAIs without evidence for live $^{26}$Al (e.g., Coki: Matzel et al., 2010) found in Comet Wild 2 may still be transported outward from the innermost disk to the comet-forming region beyond Jupiter's orbit after Jupiter has formed, with the transport being driven by the spiral arms in the remaining disk gas, which are themselves driven by Jupiter's gravitational forces (e.g., Boss and Durisen, 2005).

\section{Thermal Processing Results}

We now turn to a consideration of the thermal and chemical processing of the particles. In particular, we wish to establish whether CAIs that approach but do not hit the inner boundary would survive, and if they survive, place some limits on how much evaporation and recondensation they are likely to experience. We are also interested in how the transport of grains might locally modify the O isotopes of the nebula through the movement of inner and outer solar system ices with different O isotopic compositions. To these ends, we have calculated the thermal evaporation and condensation of grains initially consisting of a CAI-like composition core with a radius of 0.5 cm and a water ice mantle of radius 0.5 cm, for a total initial radius of 1 cm. Assuming closed system behavior, we then calculate the rate of evaporation at the ambient disk temperatures in which the grains are located (e.g., Figure 3B). Evaporation rates were calculated for cores that were assumed to have melilite-dominated mineralogies. The mineral melilite is the dominant mineral in many CAIs and is a solid solution with the more refractory end members, gehlenite and akermanite, being used here. The equilibrium mineral assemblage of a solar composition system as a function of temperature provides a useful framework for predicting when minerals are likely to evaporate or recondense. At a pressure of $10^{-5}$ bar, approximately the maximum midplane pressure near the inner boundary, in a system of solar composition, the mineral phases typically found in CAIs are stable over limited temperature ranges. Melilite, for instance, is only stable below 1430 K. Nevertheless, our calculations indicate that cores composed of either of the melilite end members would survive even if they strayed into regions of the disk with the hottest temperatures of $\sim$ 1500 K. The cores do evaporate somewhat but are largely preserved because of their size and the relatively short periods of time (a few yrs) that they remain at such temperatures.

Figure 4A shows the results of the calculations for the particle trajectory shown in Figure 3 for a gehlenite core. The initial ice shell is lost in the high temperature regions, but regained upon passage to cooler regions, although presumably the isotopic composition of the ice will have been modified by reactions with other O-bearing species. The periodic loss and gain of water ice mantles may play a direct role in the global transport of O isotopic heterogeneity in the disk. The gehlenite core experiences $\sim$ 0.15 mm of evaporation in the high temperature regions, followed by recondensation as the cores travel outwards through cooler regions. Figures 4B and 4C show the similarity of the results for a particle starting its evolution at 2.1 AU, with either a gehlenite or akermanite core. The amount of evaporation and recondensation we calculate are almost certainly overestimated and underestimated, respectively, because we do not take into account the solid solution of melilite. Dust to gas ratios higher than the solar ratio are likely to occur locally in the spiral arms during the MGU phase and should also lead to less evaporation and to the formation of thicker WL rims. While our modeling of the core evaporation/condensation is somewhat simplified, we think that our basic results are quite robust and should serve as a starting point for further investigations.

Thin Wark-Lovering (WL) rims are found on all CAIs, and their formation processes are largely unknown. For instance, the A37 CAI (Simon et al., 2011a) has a melilite core surrounded by a sequence of WL rims with compositions of spinel, hibonite, perovskite, pyroxene, and olivine, moving outward from the core. At a pressure of $10^{-5}$ bar in a system of solar composition, the equilibrium condensation temperatures of the minerals in the WL rim of A37 range from $>$ 1500 K to 1300 K. The sequence of minerals in the WL rim of A37 appears to be consistent with the short exposure of this CAI to maximum temperatures of $\sim$ 1500 K, followed by cooling to lower temperatures and the condensation of the WL rim from the disk gas, as envisioned by Simon et al. (2005), and consistent with the trajectories of many of the CAI-sized particles studied here. The midplane temperatures drop by $\sim$ 100 K over radial distances of $\sim$ 1 AU inside 3 AU. Hence we expect that WL rims should form as particles travel from near 1 AU to $\sim$ 3 AU. In addition, since there is a period of evaporation prior to formation of the WL rim, any highly refractory materials (e.g., rare-earth elements) that did not evaporate from the surfaces of the CAIs will be concentrated in the WL rims. However, Simon et al. (2005) found no evidence for an enrichment of heavy Mg isotopes in WL rims that might be expected to accompany evaporation. Large variations in the oxygen fugacity also appear to be needed to form CAI rims (Simon et al., 2005).

Stable O isotope ratios ($^{17}$O/$^{16}$O, $^{18}$O/$^{16}$O) vary by as much as 20\% in primitive meteorite components (Sakamoto, Seto, and Itoh, 2007). The dispersion in these ratios in the inner solar nebula is expected to range from $\sim$ 5\% to 15\%, based on the homogenization in a MGU disk (Boss, 2011). The initial isotopic heterogeneities may have been much larger and produced by photodissociation by ultraviolet light of CO molecules near the surface of the outer regions of the nebula (Lyons and Young, 2005). Particles traversing the inner disk are expected to encounter significant O isotope variations over distance scales of less than an AU (Boss, 2011). The O isotope variations measured in the core and rim of A37 are on the order of 5\%, and require that it condensed from an $^{16}$O-rich gas and was subsequently exposed to $^{16}$O-poor and then to $^{16}$O-rich reservoirs (Simon et al., 2011a). Such variations appear to be consistent with the expectations of our particle models and previous color homogenization models (Boss, 2011), which predict that some CAIs should have outer cores that are solar ($^{16}$O-rich), such as the compact Type A CAI EF-1 (Simon et al., 2011b), while other outer cores should be planetary ($^{16}$O-poor), as in A37 (Simon et al., 2011a).  

\section{Conclusions}

These new models imply that FU Orionis phases in MGU disks may explain several cosmochemical facts: small, refractory grains can be transported outward to great distances (Brownlee et al., 2006) and can make long, strange trips back and forth throughout the nebula (Simon et al., 2011a). In particular, the Type A Allende CAI A37 appears to have experienced a phase of evolution similar to that of a cm-sized particle in an MGU disk, resulting in the observed WL rim composition sequence and the observed range of O isotope variations (Simon et al., 2011a). If our picture is correct, we would expect WL rims of most CAIs to exhibit similar behavior, including outer cores with oxygen isotope ratios that can be either solar or planetary (Simon et al., 2011b). Finally, the expected loss of most CAIs through their accretion onto the growing protosun during FU Orionis outbursts implies that the surviving CAIs will have a relatively narrow age range, corresponding to the last one or two FU Orionis events.

\section{Acknowledgements}

We thank Denton Ebel and the referees for their advice. A.P.B.'s research was supported in part by the NASA Origins of Solar Systems Program under grant NNX09AF62G. Calculations were performed on the Carnegie Alpha Cluster, the purchase of which was supported in part by NSF MRI grant AST-9976645.

\vfill\eject
\centerline{\psfig{figure=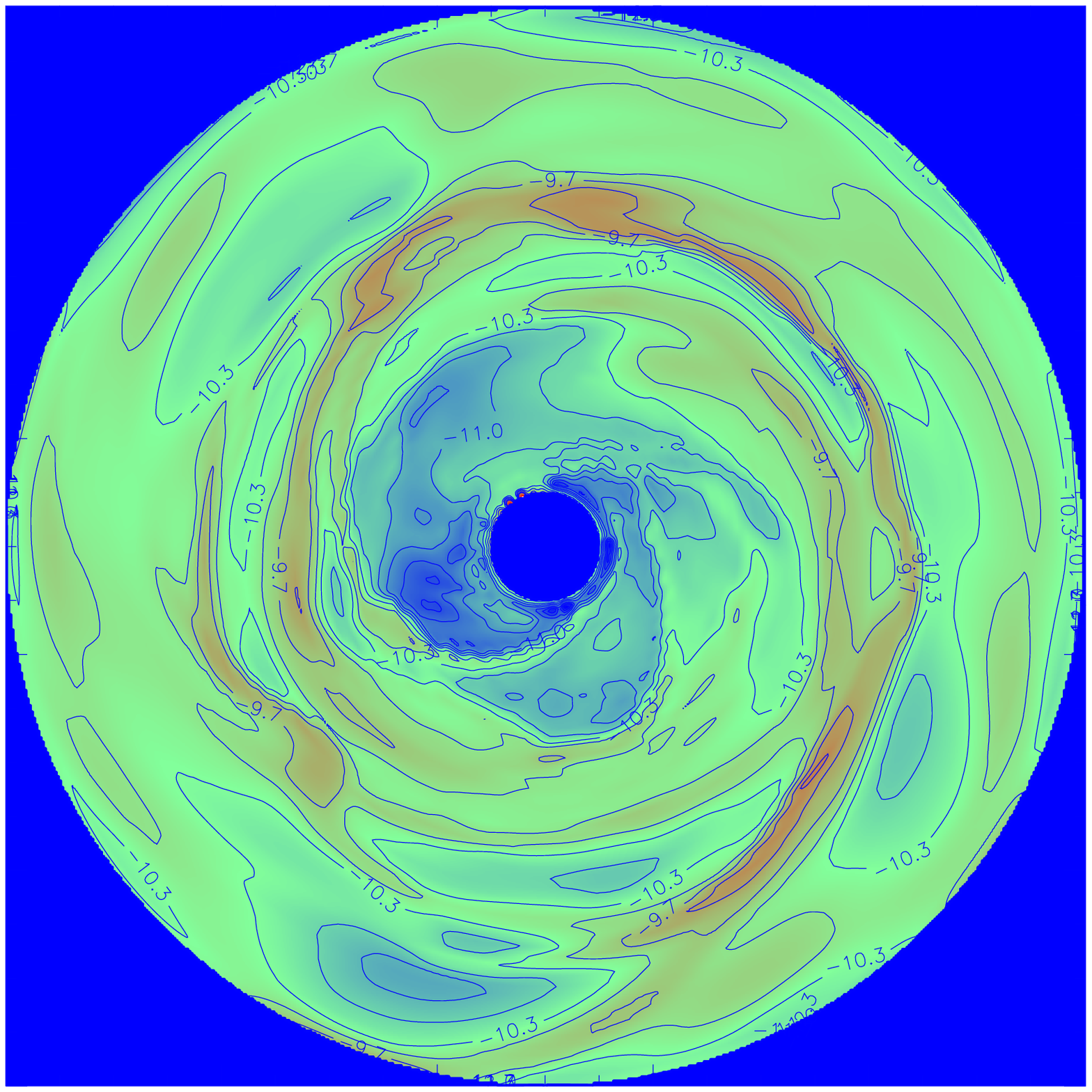,height=6.2in,width=6.2in}}
\vspace{0.2in}
Figure 1A. 
Midplane density contours (labeled in units of log g cm$^{-3}$) for the MGU disk after 105 yrs of disk evolution (A), which is also the beginning of the particle evolutions, and after 205 yrs of further disk and particle evolution (B). Midplane temperature contours (labeled in units of log K), assumed fixed throughout the evolution, are shown in C, and range from 60 K in the outer disk to 1550 K at the inner boundary. Region shown is 10 AU in radius, with an inner boundary of radius 1 AU, within which lies a solar-mass protostar.
\vspace{0.2in}

\vfill\eject
\centerline{\psfig{figure=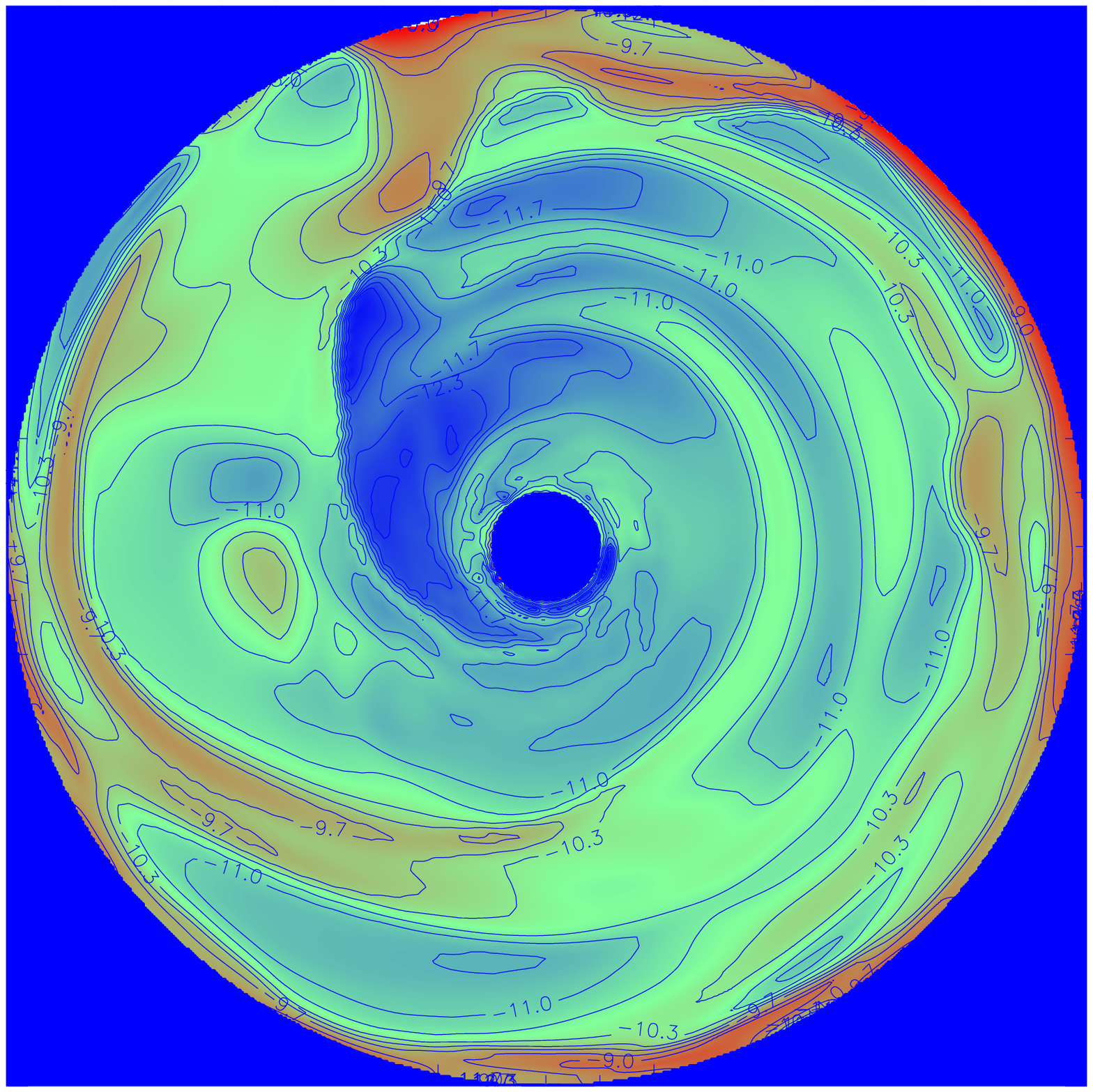,height=6.2in,width=6.2in}}
\vspace{0.2in}
Figure 1B. 
\vspace{0.2in}

\vfill\eject
\centerline{\psfig{figure=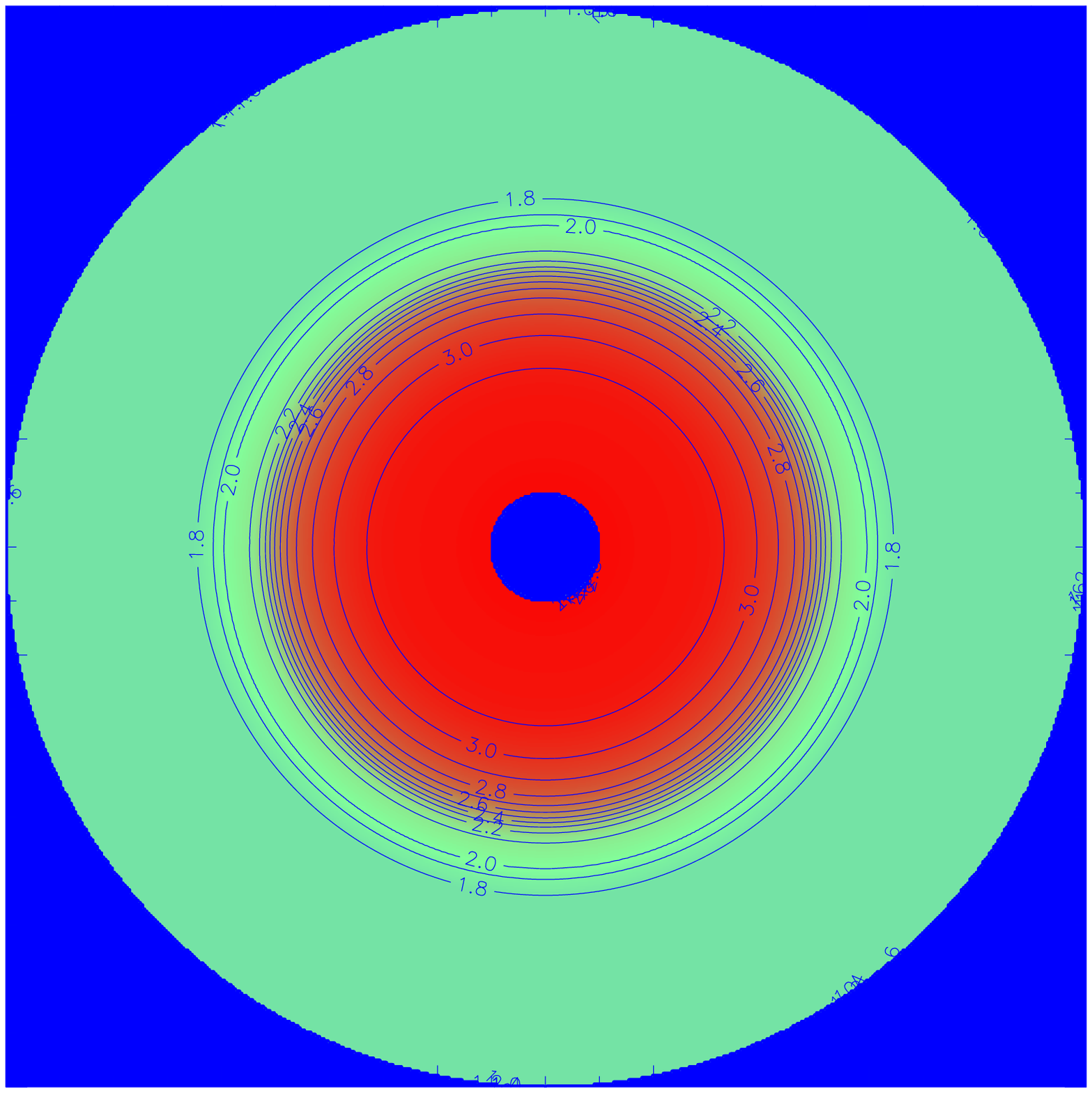,height=6.2in,width=6.2in}}
\vspace{0.2in}
Figure 1C. 
\vspace{0.2in}

\vfill\eject
\vspace{-0.7in}
\noindent
Figure 2.
Orbital evolution of a suite of 200 particles with radii of 1 cm that begin their evolutions at 9 o'clock in the disk midplane shown in Figure 1A in a band stretching from 4.5 AU to 5.5 AU. Particles that hit the inner or outer disk boundaries at 1 AU and 10 AU, respectively, are assumed to remain fixed at those distances.
\vspace{0.2in}

\vfill\eject
\centerline{\psfig{figure=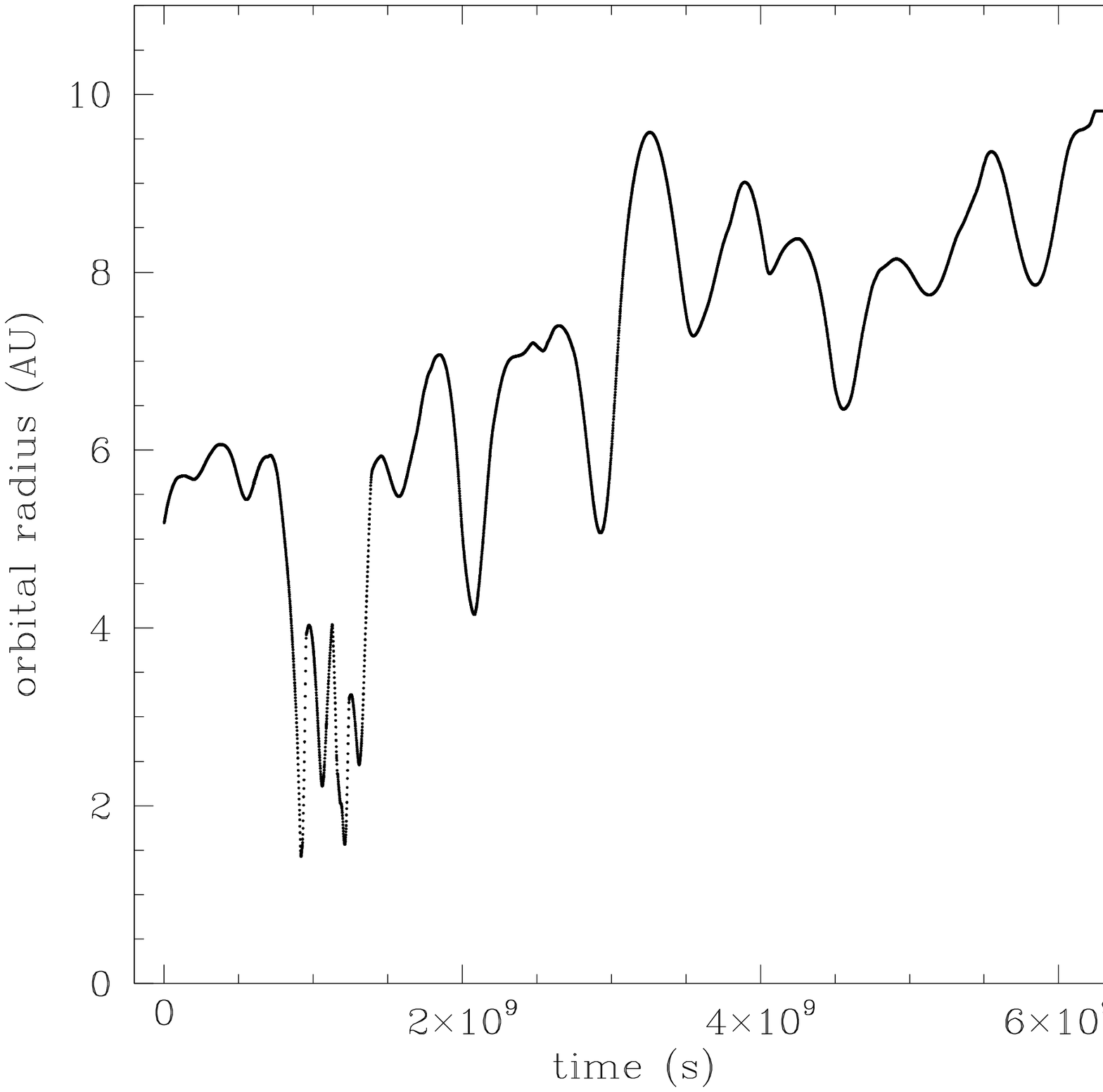,height=8.3in,width=6.2in}}
\vspace{-0.5in}
\noindent
Figure 3A.
Time evolution of the physical conditions in the disk gas experienced by a 1 cm-radius particle starting at a radius of 5.2 AU in the MGU disk shown in Figure 1A: (A) orbital radius, (B) temperature, (C) pressure, and (D) gas density.

\vspace{0.2in}

\vfill\eject
\centerline{\psfig{figure=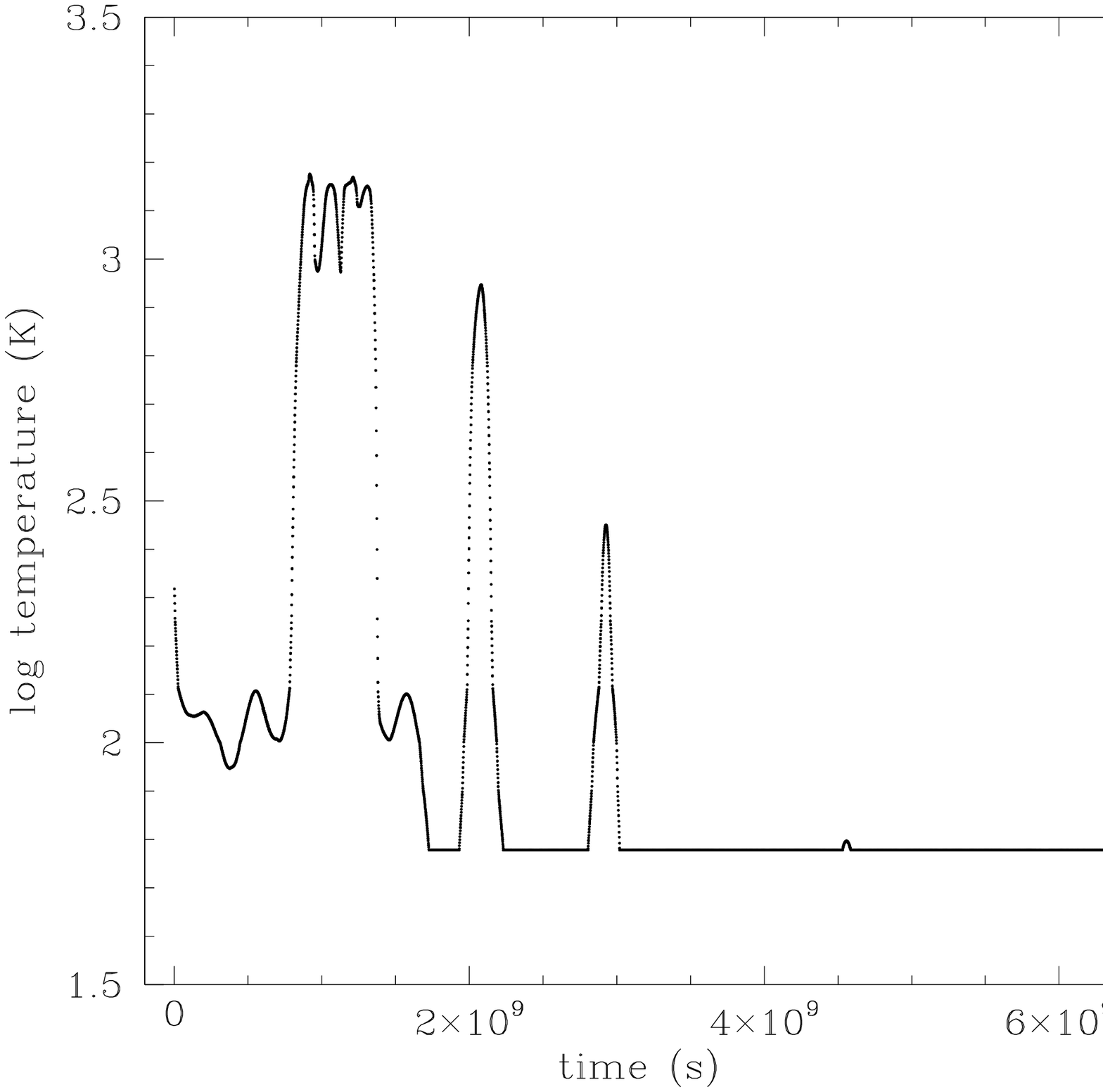,height=8.3in,width=6.2in}}
\vspace{-0.5in}
\noindent
Figure 3B.
\vspace{0.2in}

\vfill\eject
\centerline{\psfig{figure=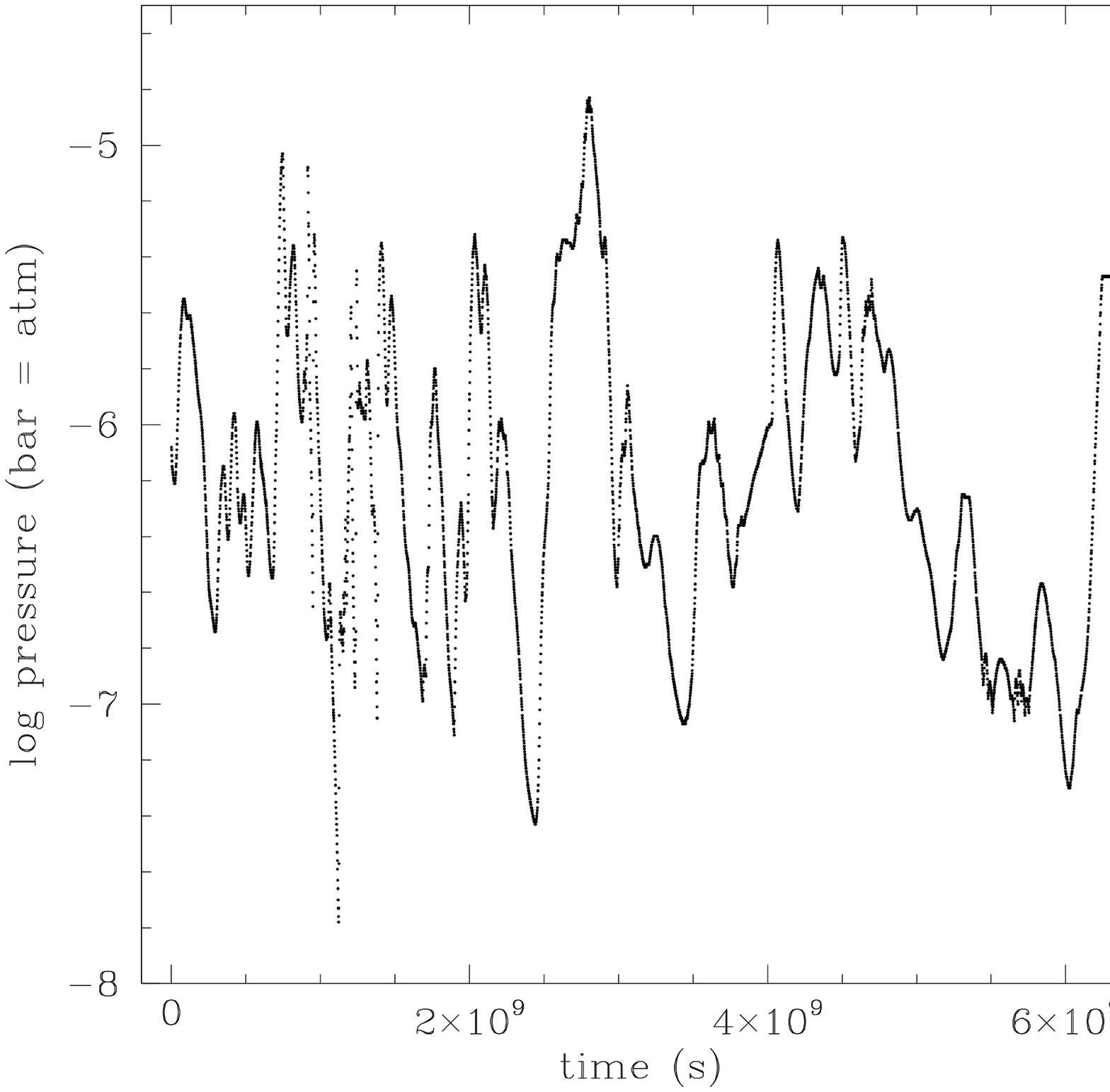,height=8.3in,width=6.2in}}
\vspace{-0.5in}
\noindent
Figure 3C.
\vspace{0.2in}

\vfill\eject
\centerline{\psfig{figure=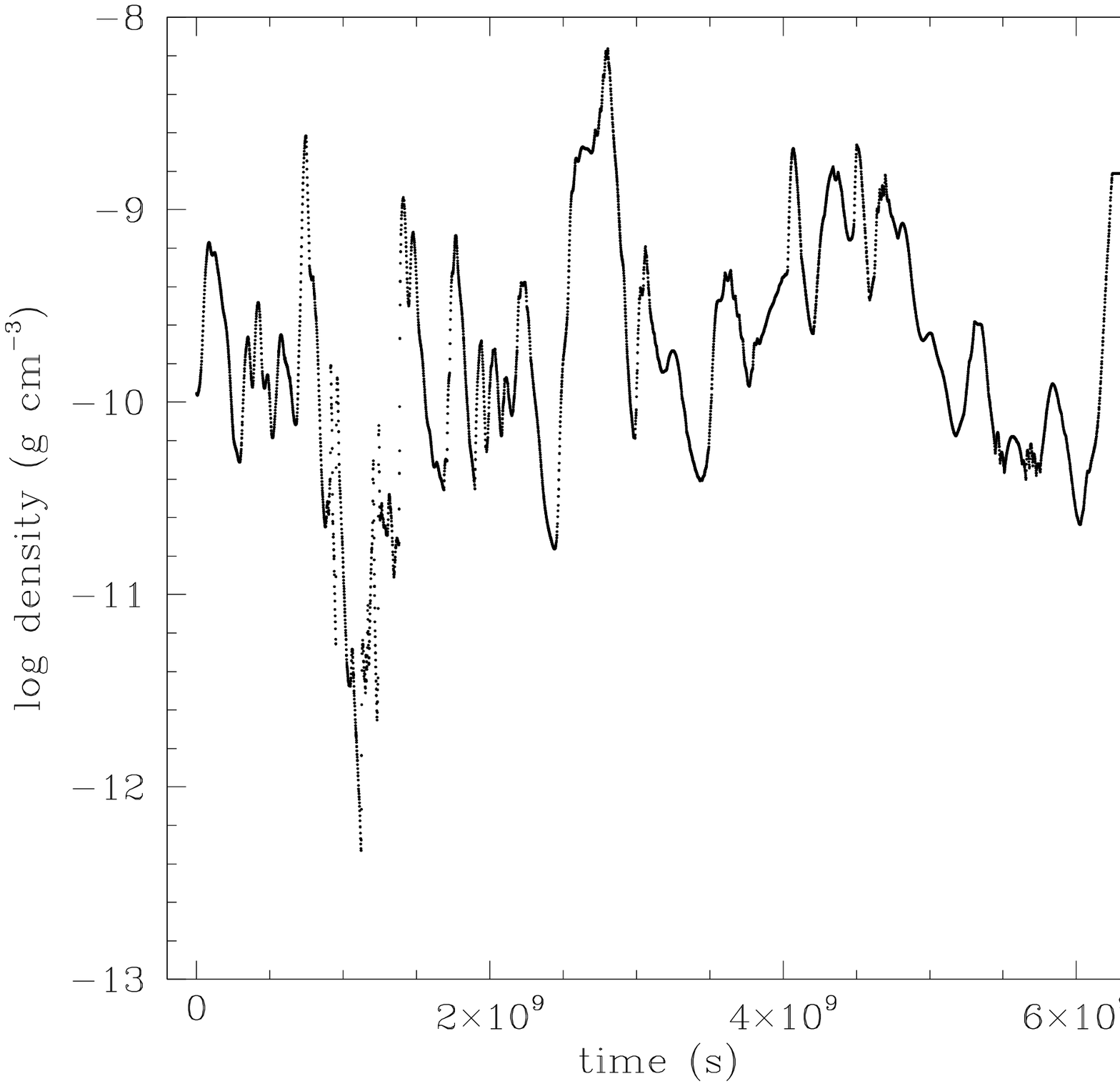,height=8.3in,width=6.2in}}
\vspace{-0.5in}
\noindent
Figure 3D.
\vspace{0.2in}

\vfill\eject
\centerline{ }
\vspace{2.0in}
\centerline{\psfig{figure=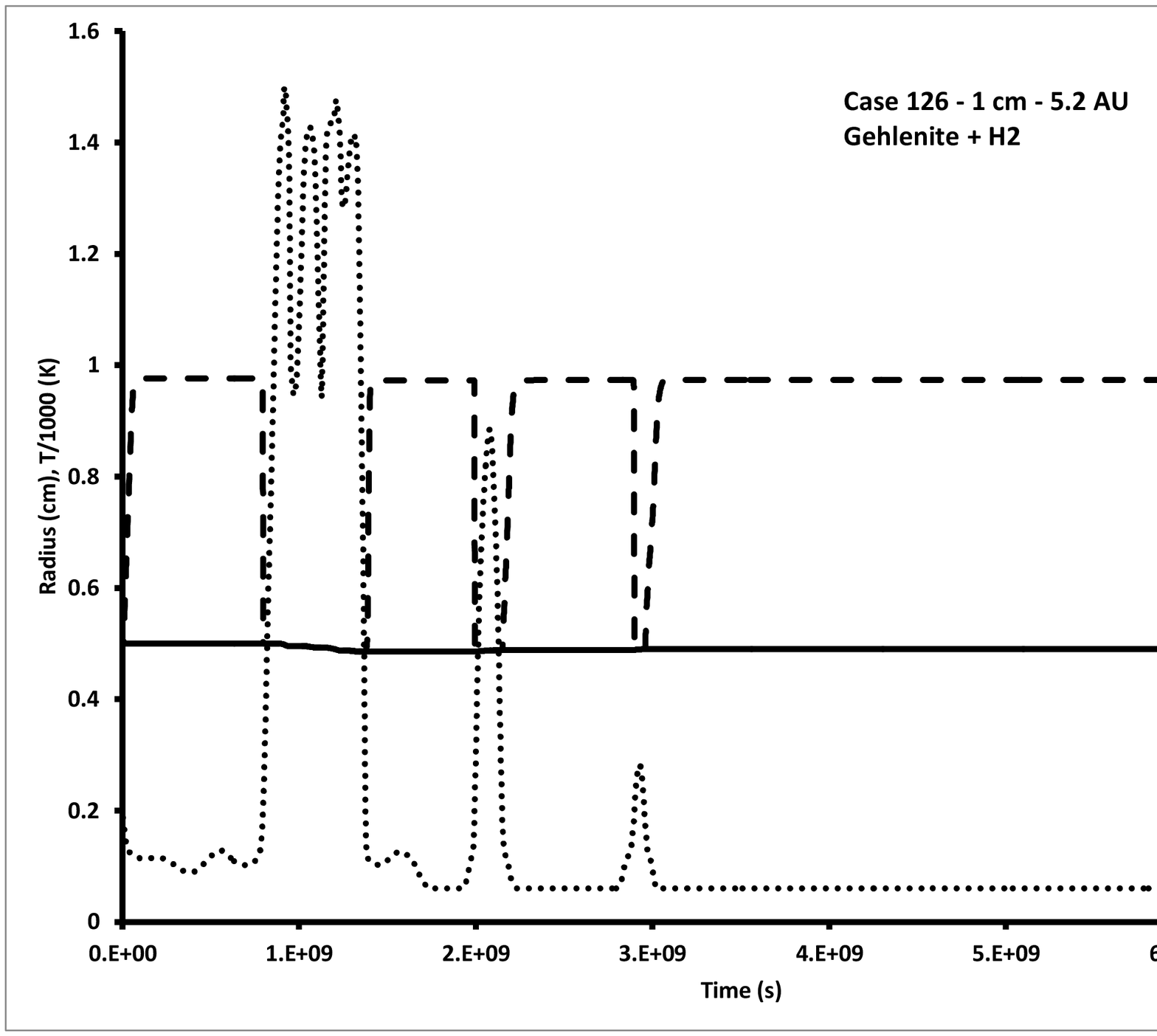,height=4.8in,width=6.2in}}
\vspace{-1.0in}
\noindent
Figure 4A.
Evolution of the 1-cm radius particle starting at 5.2 AU shown in Figure 3, assuming a gehlenite core composition (A).  Also shown are evolutions for 1-cm radius particles starting at 2.1 AU, assuming akermanite (B) or gehlenite (C) composition. In all cases the particles interact only with hydrogen gas. 
Solid lines are the radii of the particles' refractory cores, while the dashed lines are the extent of the particles' icy mantles, which are removed in hot regions and restored in cool regions in the absence of smaller particles. 
Dotted lines are the ambient disk temperatures experienced by the particles. 
Refractory cores are relatively unperturbed throughout the evolutions because of the limited time spent at high temperatures, typically less than 10 yrs at a time.
\vspace{0.2in}

\vfill\eject
\centerline{ }
\vspace{2.0in}
\centerline{\psfig{figure=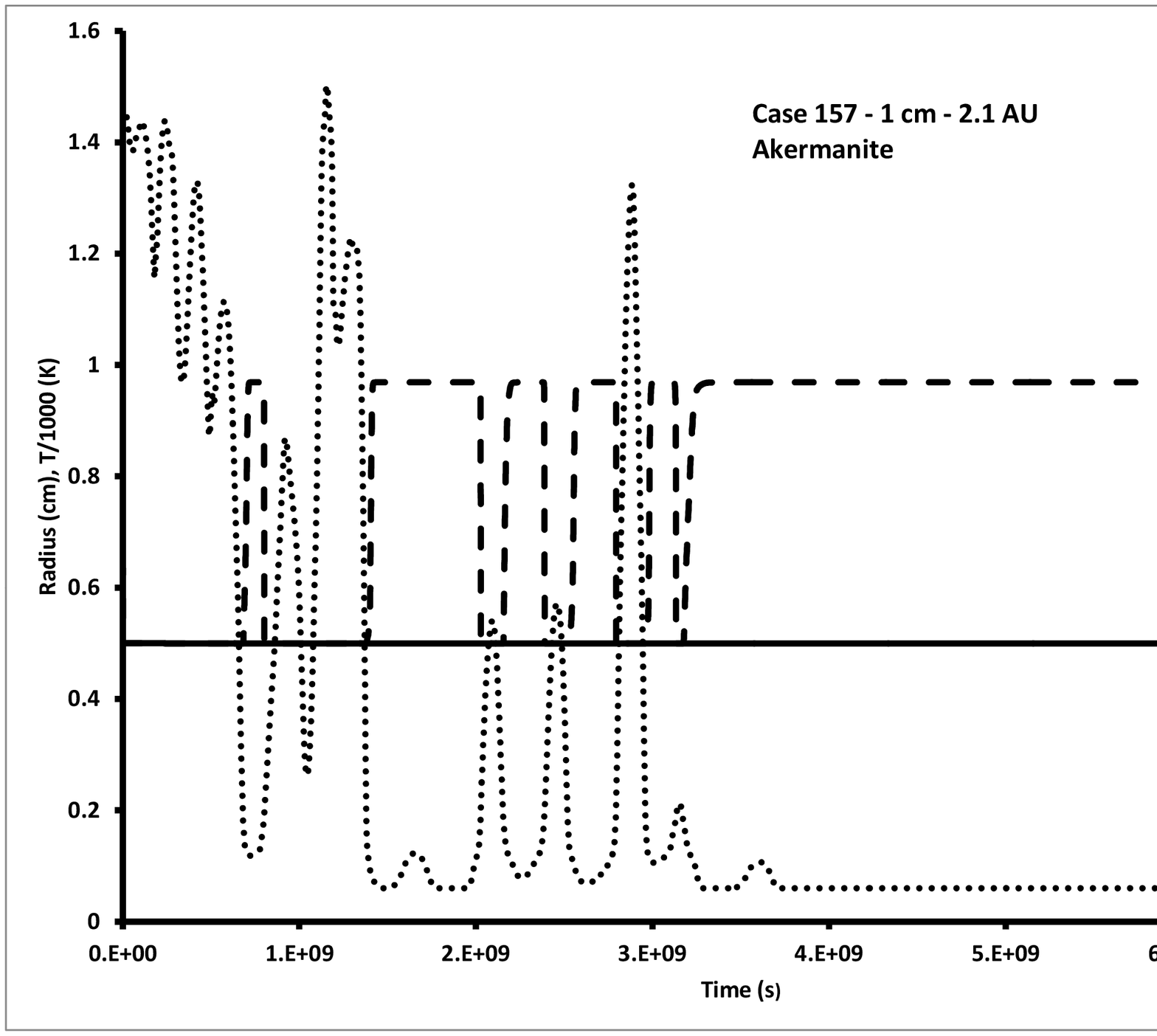,height=4.8in,width=6.2in}}
\vspace{-1.0in}
\noindent
Figure 4B.
\vspace{0.2in}

\vfill\eject
\centerline{ }
\vspace{2.0in}
\centerline{\psfig{figure=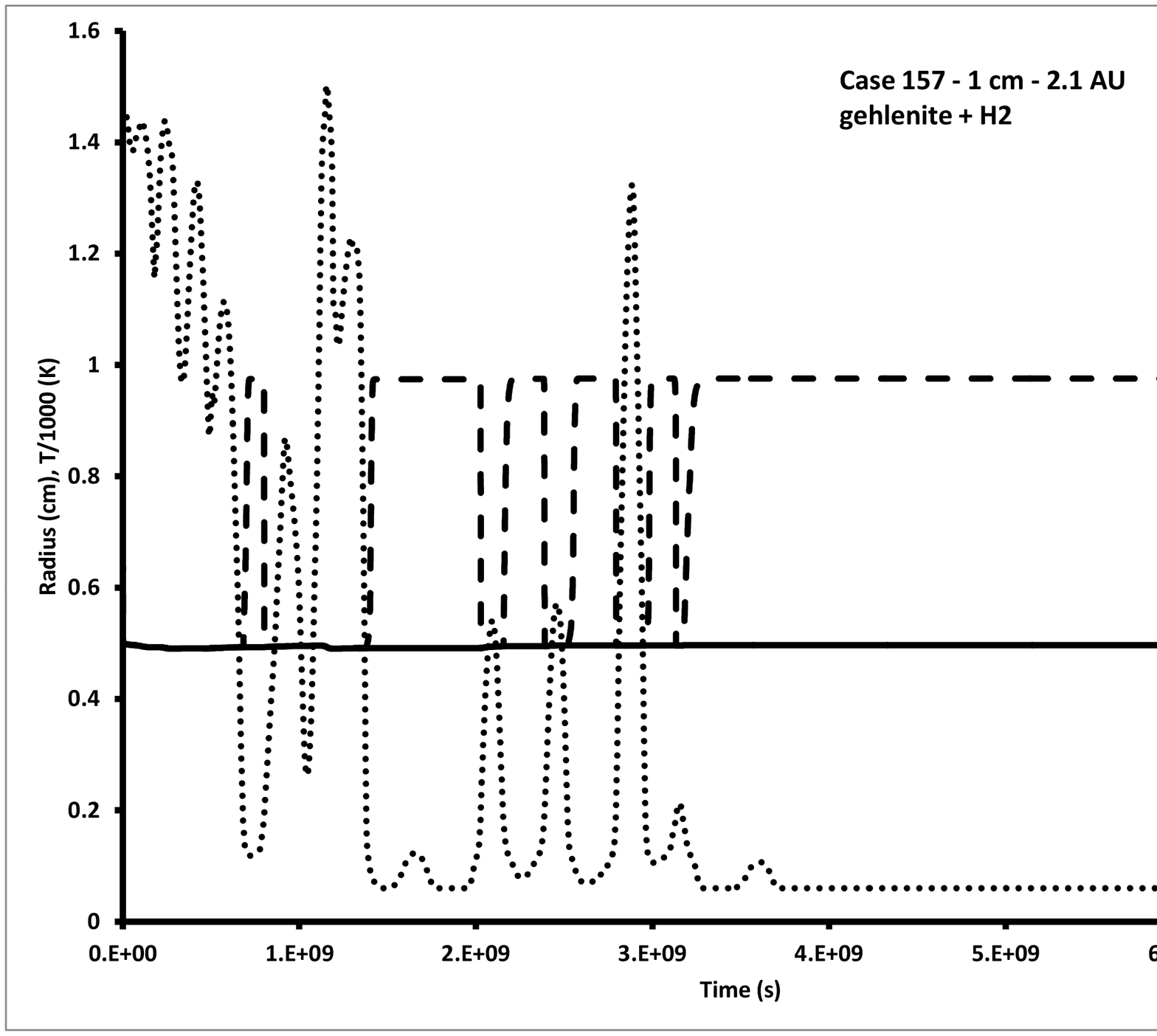,height=4.8in,width=6.2in}}
\vspace{-1.0in}
\noindent
Figure 4C.
\vspace{0.2in}

\end{document}